\documentclass[aps,prl,reprint,groupedaddress,showpacs]{revtex4-1}

\usepackage{mathrsfs}
\usepackage{graphicx}
\usepackage{epstopdf}
\usepackage{hyperref}
\begin{document}

\title{Surface Pressure of Liquid Interfaces Laden with Micron-Sized Particles}

\author{R. Mears, I. Muntz and J.~H.~J.~Thijssen}
\email[]{j.h.j.thijssen@ed.ac.uk}
\homepage[]{\\ www.ph.ed.ac.uk/people/job-thijssen}
\affiliation{SUPA School of Physics \& Astronomy, The University of Edinburgh, Edinburgh, EH9 3FD, Scotland, United Kingdom}

\date{\today}

\begin{abstract}
We consider the surface pressure of a colloid-laden liquid interface. As micron-sized particles of suitable wettability can be irreversibly bound to the liquid interface on experimental timescales, we use the canonical ensemble to derive an expression for the surface pressure of a colloid-laden interface. We use this expression to show that adsorption of particles with only hard-core interactions has a negligible effect on surface pressures from typical Langmuir-trough measurements. Moreover, we show that Langmuir-trough measurements cannot be used to extract typical interparticle potentials. Finally, we argue that the dependence of measured surface pressure on surface fraction can be explained by particle coordination number at low to intermediate particle surface fractions. At high surface fractions, where the particles are jammed and cannot easily rearrange, contact-line sliding and/or deformations of the liquid interface at the length scale of the particles play a pivotal role.
\end{abstract}

\maketitle

\section{Introduction}

Liquid interfaces laden with nano- and microparticles have received significant attention in the past few decades \cite{Binks2006}. One reason is that particle-laden liquid interfaces are model arrested systems in 2D. In addition, they have applications in materials science including (bicontinuous) Pickering emulsions \cite{Tavacoli2015,Clegg2020} and froth flotation \cite{Nguyen2006}. Moreover, studying the mechanical properties of colloid-laden interfaces provides additional insight into the mechanical properties of proteins at liquid interfaces (and \textit{vice versa}) \cite{Tcholakova2008,Jutz2010}. Proteins at liquid interfaces can play an important role in biofilms \cite{Bromley2015}, which play a role in for example healthcare \cite{Lindsay2006} and shipping \cite{Flemming2002}.

If the colloidal particles are partially wetted by both liquid/fluid phases, they can attach to the liquid interface. The detachment free energy per particle is:
\begin{equation}
    \Delta G_{\mathrm{d}} = \pi r_{\mathrm{p}}^2 \gamma_0 \left( 1 - | \cos{\theta} | \right)^2 \ ,
\label{eqn:detachment_energy}
\end{equation}
in which $r_{\mathrm{p}}$ is the particle radius, $\gamma_0$ the interfacial tension of the pristine liquid interface and $\theta$ is the three-phase contact angle \cite{Aveyard2003}. For a particle of diameter 1 $\mu$m at a water-oil interface of tension 50 mN/m, $\Delta G_{\mathrm{d}}$ can be as large as $9.5 \cdot 10^6 \ k_{\mathrm{B}}T$, where $k_{\mathrm{B}}$ is Boltzmann's constant \cite{Lide2001} and $T$ is temperature (298 K in this example). Even for a particle with a contact angle as high as 150 $^{\circ}$, $\Delta G_{\mathrm{d}} \sim 1.7 \cdot10^5 \ k_{\mathrm{B}}T$. This means that, under quiescent conditions, partially wetted micron-sized particles are irreversibly attached to liquid interfaces. This is markedly different from surfactants, as these can hop on and off the liquid interface due to thermal agitation.

The mechanical properties of particle-laden interfaces can be probed using interfacial rheology and are important for understanding the formation and stability of Pickering emulsions and bijels (bicontinuous Pickering emulsions) \cite{Thijssen2018}. Interfacial shear rheology probes the response of the interface to a shape change at constant area. Several review papers have been published on interfacial shear rheology and its applications \cite{Kragel2010, Fuller2011, Mendoza2014, Derkach2009, Miller1996, Thijssen2018}. An important dimensionless number to consider in any interfacial shear rheology experiment is the Boussinesq number:
\begin{equation}
    Bq = \frac{\eta_{\mathrm{s}}}{\eta a} \ ,
\end{equation}
in which $\eta_{\mathrm{s}}$ is the interfacial shear viscosity, $\eta$ is the viscosity of the subphase and $a$ is a dimension related to the measurement set-up \cite{Edwards1991}; for an interfacial measurement, one requires $Bq > 1$ to prevent bulk flows from dominating the measurement.

In contrast, interfacial dilational rheology measures the response of the interface to a change in area at constant shape. In a typical interfacial dilational rheology experiment, the area available to the interfacial particles $A$ is changed and the resulting change in surface pressure is measured \cite{Erni2011}. Surface pressure is a thermodynamic state variable and is defined as:
\begin{equation}{\label{eq:surface_pressure_definition}}
    \Pi = \gamma_0 - \gamma \ ,
\end{equation}
in which $\gamma$ is the apparent tension of the particle-laden interface \cite{Thijssen2018}. In a pendant-drop set-up, the tension $\gamma$ is measured by fitting the Young-Laplace equation to the measured drop profile \cite{Berry2015}. Though pendant-drop tensiometry is a popular and convenient technique, one does have to consider the potential effects of inhomogeneous particle coverage due to gravity. Moreover, the Young-Laplace equation may not apply as and when the interface becomes rigid due to compression of the particle network into a viscoelastic material \cite{Thijssen2018}. In a Langmuir-trough experiment, the interfacial tension $\gamma$ is typically measured using a Wilhelmy plate, though probes consisting of flexible beams can be used instead \cite{Gijssenbergh2018}. Notably, surface-pressure measurements using a Langmuir trough are also used as a diagnostic tool in the deposition of Langmuir-Blodget layers \cite{Deak2006}. It is worthwhile pointing out that, in a Langmuir trough experiment, there is a small shear component to the response due to a change in shape upon compression \cite{Pepicelli2017}. To apply pure dilation on a Langmuir trough setup, the development of a ``radial trough'' has recently been reported \cite{Pepicelli2017}.

One benefit of using a Langmuir trough rather than a pendant-drop set-up for measuring the mechanical properties of colloid-laden interfaces is that the gravitational force on a single particle can typically be ignored because it is negligible compared to the interfacial-tension force. This statement can be quantified using the Bond number:
\begin{equation}
    Bo = \left( \frac{r_{\mathrm{p}}}{l_{\mathrm{c}}} \right)^2 \ ,
\end{equation}
where $l_{\mathrm{c}} = \sqrt{\gamma_0 / g \Delta \rho}$ with $\Delta \rho$ the density difference between the liquids and $g$ the acceleration of gravity \cite{Vella2015}. For a 1 $\mu$m diameter sphere on a water-air interface, $Bo \sim 10^{-8} \ll 1$ confirming that gravity can be ignored. Notably, this also means flotation capillary forces, i.e.~the interparticle force due to the deformation of the liquid interface caused by particle weight, can be ignored \cite{Kralchevsky1994}. However, immersion capillary forces (for example in liquid films that are thinner than the particle diameter) or capillary forces due to contact-line undulations (for example in the case of non-spherical particles \cite{Loudet2005}) cannot be ignored \textit{a priori}. In the case of pendant-drop measurements, the gravitational force has a component parallel to the interface. This leads to particles experiencing the cumulative weight of particles above them, observed experimentally as the `keystone' mechanism \cite{Tavacoli2012}.

Previous reports have highlighted that interpreting surface-pressure measurements is challenging. For example, Du \textit{et al}.~used pendant-drop measurements to measure the detachment energy of interfacial particles \cite{Du2010}. They consider the change in total interfacial energy as particles adsorb from the bulk phase to derive an expression for the detachment energy in terms of surface pressure. Their model provides sensible values for $\Delta G_{\mathrm{d}}$ when applied to their own measurements and has been used in subsequent reports, for example Refs.~\cite{Zhang2017,Hua2016}. However, the model ignores particle-particle interactions, even though the plateau value of surface pressure is used in the analysis and it is assumed that the plateau corresponds to close packing of interfacial particles; it seems unlikely that particle-particle interactions can be ignored at close packing.

\begin{figure}
    \centering
    \includegraphics[width=0.99\linewidth]{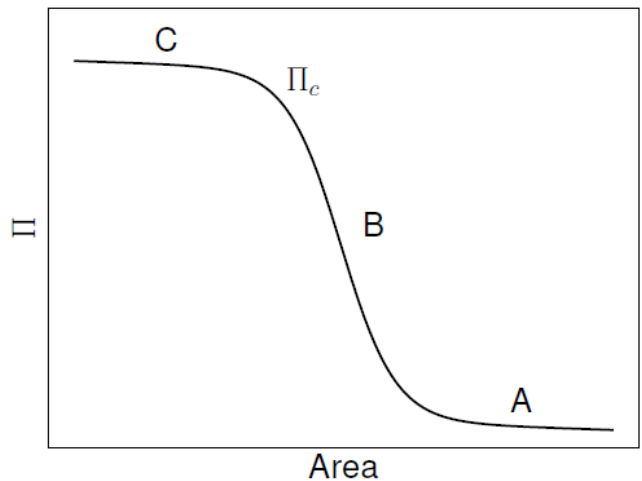}
    \caption{Schematic of surface pressure $\Pi$ vs area for a liquid interface laden with micron-sized particles. See the main text for an explanation of the critical surface pressure $\Pi_{\mathrm{c}}$ and the regions A, B and C (adapted from Ref.~\cite{Mears2020PhD}).}
    \label{fig:Figure0}
\end{figure}

Alternatively, Aveyard \textit{et al}.~used a model that only considers particle-particle interactions, i.e.~it ignores particle detachment energies, to explain the features of their measured Langmuir-trough isotherms \cite{Aveyard2000}. They identify three regions (see A, B and C in Figure \ref{fig:Figure0}) in their Langmuir isotherms. At large trough area (A), there is a slow rise of surface pressure upon compression due to long-range electrostatic interparticle repulsions. In region B, the surface pressure rises more rapidly until it levels off to a plateau at C, which the authors attribute to monolayer collapse at a critical surface pressure $\Pi_{\mathrm{c}}$ via buckling (sometimes referred to as wrinkling) rather than particle detachment. The electrostatic surface pressure model by Aveyard \textit{et al}.~successfully explains their own measurements. However, comparing this to the model by Du \textit{et al}~raises the question whether or not the particle detachment energy contributes to the surface pressure of a particle-laden interface.

In fact, there seems to be some confusion in the literature regarding the interpretation of surface pressure-area isotherms. For example, in their 2012 research paper, Fan and Striolo provide a brief overview of the debate on whether or not adsorbed particles can decrease interfacial tension (and hence increase surface pressure), noting that ``no consensus has been reached on whether the adsorbed nanoparticles affect interfacial tension''; according to their micro-Wilhelmy plate simulations, the particle detachment energy is ``not directly associated with the interfacial tension reduction'' \cite{Fan2012}. In a 2017 research paper, Zhang \textit{et al}.~note that ``despite many studies about the adsorption of particles in the interface, there appears to be no general consensus on whether simple, nonamphiphilic particles adsorbed at an interface will reduce the interfacial tension'' \cite{Zhang2017}. They continue to present a systematic, experimental study of the effect on surface pressure of silica particles of varying hydrophobicity, concluding that particles do reduce interfacial tension upon adsorption. Finally, a recent review on colloidal particles at fluid interfaces by Ballard \textit{et al}.~mentions that the ``adsorption of colloidal particles can result in a lowering of the measured interfacial tension between the two liquid phases that\ldots leads to a relation between surface tension and adsorption energy'', though they also observe that ``a significant number of experimental reports show little to no change in interfacial tension upon adsorption'' \cite{Ballard2019}. The apparent confusion regarding the interpretation of surface-pressure data for colloid-laden interfaces led us to ask ourselves: what does surface pressure mean for liquid interfaces laden with irreversibly attached colloidal particles?

Here we start by presenting a theoretical framework for the surface pressure of particles at a liquid interface that accounts for irreversible adsorption. Given the corresponding lack of chemical equilibrium between particles at the interface and those in the bulk suspension at experimentally relevant timescales, we derive an expression for surface pressure in the canonical rather than in the grand-canonical ensemble; the latter is typically used for (reversibly adsorbed) surfactants \cite{Doi2015Ch4}. We then apply our theoretical framework to previously reported surface pressure-area measurements for sterically stabilized polymer particles at a water-oil interface. Our results demonstrate that i) measured surface pressure should be negligible for low particle coverage (unless particle-particle interactions are strongly repulsive i.e.~of the order of the particle attachment energy), ii) surface-pressure measurements cannot be used to extract typical interparticle potentials in practice and iii) the shape of the isotherms at low and intermediate surface coverage can be explained in terms of particle coordination number. In addition, the magnitude of measured surface pressures implies that contact-line sliding and/or deformations of the liquid interface at the scale of the particle play a pivotal role.

\section{Theory}

For surfactants, the surface pressure is derived in the grand canonical ensemble, as the surfactant molecules at the interface are assumed to be in chemical equilibrium with the ones in the bulk subphase \cite{Doi2015Ch4}. As also pointed out by Hua \textit{et al}.~\cite{Hua2016}, this is not applicable in the case of micron-sized particles at liquid interfaces as their detachment energies are orders of magnitude larger than $k_{\mathrm{B}}T$ (equation (\ref{eqn:detachment_energy})). In other words, micron-sized particles of suitable wettability are irreversibly adsorbed, which means there is no chemical equilibrium between the colloidal particles at the interface and those in the bulk (sub)phase at experimental timescales. Hence, we proceed below to derive an expression for the surface pressure of a colloid-laden liquid interface in the canonical ensemble.

\subsection{Surface pressure in the canonical ensemble}

We first consider the osmotic pressure $\Pi_{\rm 3D}$ of a suspension of a solute in a solvent \cite{Doi2015Ch2}. The osmotic pressure is the force per unit area that is required to hold in place a semi-permeable membrane between a volume of suspension and a volume of solvent. The surface pressure can be interpreted as the force per unit length that is required to hold in place a semi-permeable barrier between an area of particle-laden liquid interface and an area of pristine liquid interface. Hence, surface pressure is the 2D equivalent of osmotic pressure. In the canonical ensemble, the osmotic pressure can be written as:
\begin{equation}{\label{eq:can_osmotic_pressure}}
	\begin{array}{rcl}
		\Pi_{\rm 3D} \left( \varphi \right) & = & f_{\rm 3D} \left( 0 \right) - f_{\rm 3D} \left( \varphi \right) + \varphi \frac{\partial  f_{\rm 3D}}{\partial \varphi} \\
		\vspace{0.5mm} \\
		& = & f_{\rm 3D} \left( 0 \right) - f_{\rm 3D} \left( n \right) + n \frac{\partial  f_{\rm 3D}}{\partial n} \ ,
	\end{array}
\end{equation}
in which $\varphi$ is the volume fraction of solute, $f_{\rm 3D}$ is the free energy per unit volume, and $n = \varphi / v_{\rm p}$ is the number density in 3D, with $v_{\rm p}$ the volume of a solute particle. For the surface pressure $\Pi$, we can write the 2D equivalent of equation (\ref{eq:can_osmotic_pressure}):
\begin{equation}{\label{eq:can_surface_pressure}}
	\Pi \left( \rho \right) = \gamma_0 - f \left( \rho \right) + \rho \mu \ .
\end{equation}
Here $\rho$ is the particle number density in 2D (equivalent to $n$ in 3D), $f$ is the free energy per unit area, $f \left( 0 \right) = \gamma_0$ (at $\rho = 0$ the free energy per unit area is the interfacial tension of the pristine interface), and $\mu = \left( \partial  f / \partial \rho \right)_{A,T}$ is the chemical potential; see the Supplementary Information for a derivation starting from the  (canonical) free energy $F$ of the particle-laden interface.

\subsection{Example 1: particles with hard-core interactions}

We consider the situation in which $N$ particles from the bulk have attached to the liquid interface of fixed area $A$ at fixed temperature $T$. We assume that the particles do not interact with each other, apart from that they cannot overlap, and we ignore their entropy (see equation (\ref{eq:surface_pressure_virial})). The free energy of the liquid interface before particle attachment is:
\begin{equation}{\label{eq:free_energy_liquid_interface}}
F_{\mathrm{li}} = \gamma_0 A \ .
\end{equation}
Each particle attaching to the interface lowers the interfacial free energy by an amount $\Delta G_{\mathrm{d}}$ (equation (\ref{eqn:detachment_energy})):
\begin{equation}{\label{eq:free_energy_liquid_interface_particles}}
F_{\mathrm{lip}} = \gamma_0 A - N \Delta G_{\mathrm{d}} \ .
\end{equation}

We can now calculate the free energy per unit area,
\begin{equation}{\label{eq:free_energy_density_liquid_interface}}
f_{\mathrm{lip}} = \frac{F_{\mathrm{lip}}}{A} = \gamma_0 - \rho \Delta G_{\mathrm{d}} \ ,
\end{equation}
and the chemical potential,
\begin{equation}{\label{eq:chemical_potential_liquid_interface}}
\mu = \left( \frac{\partial f}{\partial \rho} \right)_{A,T} = -\Delta G_{\mathrm{d}} \ ,
\end{equation}
resulting in the following for the surface pressure:
\begin{equation}{\label{eq:surface_pressure_liquid_interface}}
\begin{array}{rcl}
\Pi & = & \gamma_0 - f + \rho \mu \\
\vspace{0.5mm} \\
& = & \gamma_0 - \gamma_0 + \rho \Delta G_{\mathrm{d}} - \rho \Delta G_{\mathrm{d}} = 0 \ .
\end{array}
\end{equation}
This result aligns with measurements at low surface fractions, where the particles are expected not to interact (see Figure \ref{fig:Figure1}).

As mentioned above, we have neglected the contribution of the entropy of the colloids to the surface pressure, as well as any entropy change due to structuring of the molecules of the dispersing medium around the colloids. Comparing to the equivalent equation for osmotic pressure in dilute suspensions of solutes in 3D \cite{Doi2015Ch2}:
\begin{equation}{\label{eq:osmotic_pressure_dilute}}
\Pi_{\mathrm{o}} = \frac{\varphi k_{\mathrm{B}} T}{v_{\mathrm{p}}} \ ,
\end{equation}
we can write the entropy-contribution to surface pressure at low surface fractions as:
\begin{equation}{\label{eq:surface_pressure_virial}}
\Pi_{\mathrm{S}} = \frac{k_{\mathrm{B}} T}{a_{\mathrm{p}}} \phi \ ,
\end{equation}
in which $a_{\mathrm{p}} = \pi r_{\mathrm{p}}^2$ is the cross-sectional area per particle. Equation (\ref{eq:surface_pressure_virial}) predicts that $\Pi_{\mathrm{S}} \sim 10^{-6}$ mN/m for $r_{\mathrm{p}} = 1$ $\mu$m, which is negligible compared to typical values of measured interfacial tension $\gamma_0$ and surface pressure $\Pi$ (see Figure \ref{fig:Figure1}). In fact, according to equation (\ref{eq:surface_pressure_virial}), $\Pi_{\mathrm{S}}$ is of order 1 mN/m only if $r_{\mathrm{p}}$ is of order 1 nm or smaller, which is closer to the lengthscale of a typical surfactant molecule. These considerations imply that, for micron-sized colloidal particles at liquid interfaces, the contribution of their entropy to the surface pressure is negligible.

\subsection{Example 2: beyond hard-core interactions}

We now consider the case of particles that have interactions in addition to hard-core repulsion. Following on from equation (\ref{eq:free_energy_liquid_interface_particles}), additional interactions between the particles lead to an additional term in the free energy:
\begin{equation}{\label{eq:interact_free_energy_total}}
F_{\mathrm{lip}} = \gamma_0 A - N \Delta G_{\mathrm{d}} + N \bar{f}_{\mathrm{p}} \ ,
\end{equation}
where $\bar{f}_{\mathrm{p}}$ is the (average) free energy per particle due to particles interacting. Note that $\bar{f}_{\mathrm{p}}$ includes the effects of pair potentials between interfacial particles but also, for example, the additional energy barrier to particle attachment caused by particles already attached to the interface. If we imagine a compression experiment in a Langmuir-trough that starts with a relatively low surface coverage, the latter contribution can be ignored.

We can now calculate the free energy per unit area,
\begin{equation}{\label{eq:interact_free_energy_density_liquid_interface}}
f_{\mathrm{lip}} = \frac{F_{\mathrm{lip}}}{A} = \gamma_0 - \rho \Delta G_{\mathrm{d}} + \rho \bar{f}_{\mathrm{p}} \ ,
\end{equation}
and the chemical potential,
\begin{equation}{\label{eq:interact_chemical_potential_liquid_interface}}
\begin{array}{rcl}
\mu & = & \left( \frac{\partial f_{\mathrm{lip}}}{\partial \rho} \right)_{A,T} \\
\vspace{0.5mm} \\
& = & -\Delta G_{\mathrm{d}} + \bar{f}_{\mathrm{p}} + \rho \left( \frac{\partial \bar{f}_{\mathrm{p}}}{\partial \rho} \right)_{A,T} \ ,
\end{array}
\end{equation}
yielding the following for the surface pressure:
\begin{equation}{\label{eq:interact_surface_pressure_init}}
\begin{array}{rcl}
\Pi & = & \gamma_0 - f + \rho \mu \\
\vspace{0.5mm} \\
& = & \rho^2 \left( \frac{\partial \bar{f}_{\mathrm{p}}}{\partial \rho} \right)_{A,T} \ .
\end{array}
\end{equation}

\subsection{Comparison to Langmuir-trough experiments}

The expression for surface pressure $\Pi$ in equation (\ref{eq:interact_surface_pressure_init}) involves partial derivatives at fixed area $A$ and temperature $T$. However, typical Langmuir-trough experiments on micron-sized particles at liquid interfaces are performed at constant temperature $T$ and number of interfacial particles $N$. Here, we derive an expression for surface pressure $\Pi$, at fixed $T$ and $N$, as a function of surface fraction $\phi$.

We start with the differential of the canonical free energy $F$ in 2D,
\begin{equation}\label{eqn:diff_can_free_energy_2D}
	\mathrm{d}F = \gamma \mathrm{d}A - S \mathrm{d}T + \mu \mathrm{d}N \ ,
\end{equation}
in which $S$ is the entropy of the 2D system (see SI) \cite{Doi2015Ch4,Kittel1998}. Inserting equation (\ref{eq:interact_free_energy_total}) results in:
\begin{equation}{\label{eq:tension_differential}}
\begin{array}{rcl}
\gamma & = & \left( \frac{\partial F_{\mathrm{lip}}}{\partial A} \right)_{N,T} \\
\vspace{0.5mm} \\
& = & \gamma_0 + \left( \frac{\partial F_{\mathrm{p}}}{\partial A} \right)_{N,T} \ ,
\end{array}
\end{equation}
with $F_{\mathrm{p}}$ the total free energy due to particles interacting. For the surface pressure $\Pi$ (equation (\ref{eq:surface_pressure_definition})), we can then write:
\begin{equation}{\label{eq:surface_pressure_particles}}
\Pi = -\left( \frac{\partial F_{\mathrm{p}}}{\partial A} \right)_{N,T} \ .
\end{equation}

As the number of interfacial particles $N$ is kept fixed, we can rewrite equation (\ref{eq:surface_pressure_particles}) as:
\begin{equation}{\label{eq:free_energy_particle_pi}}
\left( \frac{\partial \left( F_{\mathrm{p}} / N \right)   }{\partial \left( A/N \right)} \right)_{N,T} = \left( \frac{\partial \bar{f}_{\mathrm{p}}}{\partial a} \right)_{N,T} = -\Pi \ ,
\end{equation}
in which $a$ is the (average) interfacial area per particle. As
\begin{equation}{\label{eq:from_a_to_phi}}
\phi = \frac{N a_{\mathrm{p}}}{A} = \frac{a_{\mathrm{p}}}{a} \ ,
\end{equation}
we can write equation (\ref{eq:free_energy_particle_pi}) as:
\begin{equation}{\label{eq:free_energy_particle_phi_diff}}
\left( \frac{\partial \bar{f}_{\mathrm{p}}}{\partial \phi}  \right)_{N,T} = \frac{a_{\mathrm{p}} \Pi}{\phi^2} \ .
\end{equation}

To obtain the average free energy per particle due to particles interacting, we can integrate equation (\ref{eq:free_energy_particle_phi_diff}):
\begin{equation}{\label{eq:free_energy_particle_phi}}
\bar{f}_{\mathrm{p}} \left( \phi \right) = a_{\mathrm{p}} \int_0^{\phi} \frac{\Pi}{\phi'^{2}} \mathrm{d} \phi' \ .
\end{equation}
Note that equation (\ref{eq:free_energy_particle_phi}) suggests that $\bar{f}_{\mathrm{p}}$ can be obtained via numerical integration of surface-pressure measurements at constant $N$ and $T$, for example Langmuir-trough measurements.

With a few additional assumptions, we can extract interparticle potentials from Langmuir-trough measurements. First, as explained just below equation (\ref{eq:surface_pressure_virial}), we assume that the contribution of the entropy of the particles to the surface pressure is negligible, which means equation (\ref{eq:free_energy_particle_phi_diff}) can be written as:
\begin{equation}{\label{eq:surface_pressure_energy_particle}}
\Pi \approx \frac{\phi^2}{a_{\mathrm{p}}} \left( \frac{\partial \bar{u}_{\mathrm{p}}}{\partial \phi} \right)_{N,T} \ ,
\end{equation}
in which $\bar{u}_{\mathrm{p}}$ is the internal interaction energy per particle. Secondly, we will assume that particles only interact with, on average, $\bar{z} \left( \phi \right)$ nearest neighbours via an interparticle potential $\bar{\epsilon}_{\mathrm{pp}} \left( \phi \right)$. In that case,
\begin{equation}{\label{eq:surface_pressure_interparticle_potential}}
\Pi \approx \frac{\phi^2}{2 a_{\mathrm{p}}} \left( \frac{\partial \left( \bar{z} \left( \phi \right) \bar{\epsilon}_{\mathrm{pp}}\left( \phi \right) \right)}{\partial \phi} \right)_{N,T} \ ,
\end{equation}
where the division by $2$ prevents double-counting of particle-particle pairs. Integrating equation (\ref{eq:surface_pressure_interparticle_potential}), we finally arrive at:
\begin{equation}{\label{eq:interparticle_potential_langmuir_trough}}
\bar{\epsilon}_{\mathrm{pp}} \left( \phi \right) \approx \frac{2 \bar{f}_{\mathrm{p}} \left( \phi \right)}{\bar{z} \left( \phi \right)} \ ,
\end{equation}
for which $\bar{f}_{\mathrm{p}}$ can be obtained from equation (\ref{eq:free_energy_particle_phi}). Notably, equation (\ref{eq:interparticle_potential_langmuir_trough}) provides a route, in theory, to determining interparticle potentials from Langmuir-trough measurements via numerical integration of experimental data using equation (\ref{eq:free_energy_particle_phi}) \emph{if} $z \left( \phi \right)$ is known.

\subsection{Comparison with model by Du \textit{et al}.}

We can also compare our theoretical results to what we find with a straightforward revision of the model set out by Du \textit{et al.}~\cite{Du2010}. Here we consider a section of the interface with area $A$ and number of particles $N$. The energy associated with this setup is given as,
\begin{equation}{\label{eq:du_start_energy}}
    E(A) = \gamma_0A + N\bar{u}_{\mathrm{p}} - N\Delta G_{\mathrm{d}} \ .
\end{equation}
Equation (\ref{eq:du_start_energy}) is similar to equation (\ref{eq:interact_free_energy_total}) but, following Du \emph{et al}, we use energy rather than free energy here.

Consider now increasing the interfacial area by a small amount $\mathrm{d}A$, while maintaining a fixed number of particles, and allowing $\bar{u}_p = \bar{u}_p(A)$. We can then write the associated energy as:
\begin{equation}{\label{eq:du_da_energy}}
    \begin{array}{rcl}
        E(A+\mathrm{d}A) & = & \gamma_0A+\gamma_0\mathrm{d}A \\
        \vspace{0.1mm} \\
         & + & N\bar{u}_p(A+\mathrm{d}A) - N\Delta G_{\mathrm{d}} \ .
    \end{array}
\end{equation}
We can use this to find the change in energy $\mathrm{d}E$ upon the change in area $\mathrm{d}A$,
\begin{equation}{\label{eq:du_inf_energy}}
    \begin{array}{rcl}
        \mathrm{d}E & = & E(A+\mathrm{d}A) - E(A) \\
        \vspace{0.5mm} \\
         & = & \gamma_0\mathrm{d}A + N(\bar{u}_p(A+\mathrm{d}A) - \bar{u}_p(A)) \ ,
    \end{array}
\end{equation}
which, upon expansion of $\bar{u}_p(A+\mathrm{d}A)$ to first order in $\mathrm{d}A$, can be written as
\begin{equation}{\label{eq:du_inf_energy_full}}
    \mathrm{d}E = \gamma_0\mathrm{d}A + N\left(\frac{\partial\bar{u}_p}{\partial A}\right)_{N, T} \mathrm{d}A \ .
\end{equation}

Following Du \textit{et al}., we then write the interfacial tension of the particle laden surface as
\begin{equation}{\label{eq:du_int_tension}}
    \gamma = \left( \frac{\partial E}{\partial A} \right)_{N, T} \ .
\end{equation}
Combining equations (\ref{eq:du_inf_energy_full}) and (\ref{eq:du_int_tension}) leads to
\begin{equation}
    \gamma = \gamma_0 + N\left(\frac{\partial\bar{u}_p}{\partial A}\right)_{N, T} \ ,
\end{equation}
or equivalently (see equation (\ref{eq:surface_pressure_definition})):
\begin{equation}
    \Pi = -N\left(\frac{\partial\bar{u}_p}{\partial A}\right)_{N, T} \ .
\end{equation}
A change of variable from $A$ to $\rho = N / A$, and then to $\phi = a_{\mathrm{p}} \rho$, results in
\begin{equation}{\label{eq:pi_from_du}}
    \Pi = \rho^2\left(\frac{\partial\bar{u}_p}{\partial \rho}\right)_{N, T} = \frac{\phi^2}{a_{\mathrm{p}}} \left( \frac{\partial \bar{u}_{\mathrm{p}}}{\partial \phi} \right)_{N,T} \ ,
\end{equation}
which is equivalent to equation (\ref{eq:surface_pressure_energy_particle}), and similar to equation (\ref{eq:interact_surface_pressure_init}), if entropy is ignored.

\subsection{Interparticle potentials}

For poly(methyl methacrylate) (PMMA) particles stabilized by poly(12-hydroxystearic acid) (PHSA) at a water-alkane interface, as considered below, Muntz \emph{et al}.~have recently measured the interfacial pair potential $\bar{\epsilon}_{\mathrm{pp}}$ using fluorescence microscopy and optical tweezers. At low $r$,
\begin{equation}{\label{eq:muntz_u}}
\bar{\epsilon}_{\mathrm{pp}} \left( r \right) = \left( \frac{\alpha}{r} \right) \mathrm{e}^{-\kappa r} \ 
\end{equation}
provides a decent fit to the measuresments \cite{Muntz2018}. Here $r$ is the distance between particles, $\alpha$ is a prefactor with value $4.1 \cdot 10^3 \ k_{\mathrm{B}} T \mathrm{\mu m}$ and $\kappa$ is the inverse Debye screening length with value $0.35 \ \mathrm{\mu m}^{-1}$.

If we assume that the interfacial particles are arranged in a hexagonal pattern and only interact with their $z$ nearest neighbours, we can write:
\begin{equation}{\label{eq:hex_u}}
\bar{u}_{\mathrm{p}} \left( r \right) = \frac{z}{2} \left( \frac{\alpha}{r} \right) \mathrm{e}^{-\kappa r} \ .
\end{equation}
For the contribution of interparticle interactions to the surface pressure, following equation (\ref{eq:pi_from_du}), we can then write:
\begin{equation}{\label{eq:hex_surface_pressure_ubar}}
\begin{array}{rcl}
\Pi_{U} & = & \rho^2 \left( \frac{\partial \bar{u}_{\mathrm{p}}}{\partial \rho} \right)_{N,T} \\
\vspace{0.5mm} \\
& = & \rho^2 \left( \frac{\partial r}{\partial \rho} \right)_{N, T} \left( \frac{\partial \bar{u}_{\mathrm{p}}}{\partial r} \right)_{N,T} \\
\vspace{0.5mm} \\
& = & \frac{z}{2 \sqrt{3} r^2} \alpha \mathrm{e}^{-\kappa r} \left( \kappa + \frac{1}{r} \right) \ ,
\end{array}
\end{equation}
where we have used:
\begin{equation}{\label{eq:hex_rho_vs_r}}
\rho = \frac{2}{r^2 \sqrt{3}} \ ,
\end{equation}
for a hexagonal pattern of interfacial particles.

Note that equation (\ref{eq:hex_surface_pressure_ubar}) predicts that repulsive interactions between interfacial particles contribute to a higher surface pressure, which is in line with previous reports \cite{Hua2016}. However, even at $\phi = 0.9$ i.e.~$r \approx 2.008 r_{\mathrm{p}}$, $\Pi_U \sim 0.003$ mN/m for $z = 6$, $r_{\mathrm{p}} = 1$ $\mu$m and $T = 298$ K. Hence, we would expect that these particles at a liquid interface do not lead to a substantial surface pressure until they start percolating, at which point contact forces should be considered. Given typical errors in surface-pressure measurements, this also means that extracting this colloidal pair potential from Langmuir-trough measurements does not seem feasible.

At this point, one might argue that the surface pressure could be substantially higher for charged particles at a water-oil interface. Hence, we apply a similar analysis to a system of 3.1 $\mu$m diameter polystyrene particles at a water-decane interface \cite{Masschaele2010}. Masschaele \textit{et al}.~compare the following interparticle potential:
\begin{equation}{\label{eq:interparticle_potential_masschaele}}
    \bar{\epsilon}_{\mathrm{pp}} = k_{\mathrm{B}} T \left( \frac{a_1}{3 r} \mathrm{e}^{-\kappa r} + \frac{a_2}{r^3} \right) \ ,
\end{equation}
to experimental data using $a_1 \sim 235$ m and $\kappa^{-1} = 300$ nm; the experimentally determined upper bound of $a_2$ is of order $10^{-13}$ m$^3$. For ease of comparison, we re-write equation (\ref{eq:interparticle_potential_masschaele}) as:
\begin{equation}{\label{eq:interparticle_potential_masschaele_scaled}}
    \begin{array}{rcl}
        \bar{\epsilon}_{\mathrm{pp}} & = & k_{\mathrm{B}} T \left( \frac{a_1 / r_{\mathrm{p}}}{3r/r_{\mathrm{p}}} \mathrm{e}^{-\kappa r_{\mathrm{p}} r / r_{\mathrm{p}}} + \frac{a_2/r_{\mathrm{p}}^3}{r^3/r_{\mathrm{p}}^3} \right) \\
        \vspace{1.0mm} \\
        & = & k_{\mathrm{B}} T \left( \frac{a_{1\mathrm{s}}}{3x} \mathrm{e}^{-\kappa r_{\mathrm{p}}x} + \frac{a_{2\mathrm{s}}}{x^3} \right) \ ,
    \end{array}
\end{equation}
where $x = r/r_{\mathrm{p}}$, $a_{1\mathrm{s}} = a_1 / r_{\mathrm{p}}$ and $a_{2\mathrm{s}} = a_2 / r_{\mathrm{p}}^3$. Next, we take the derivative of equation (\ref{eq:interparticle_potential_masschaele_scaled}) with respect to $r$ (and multiply the result by $-1$), in order to obtain the interparticle force:
\begin{equation}{\label{eq:interparticle_force_masschaele}}
    f_{\mathrm{pp}} = \frac{k_{\mathrm{B}} T}{r_{\mathrm{p}}} \frac{1}{x} \left( \frac{a_{1\mathrm{s}}}{3} \mathrm{e}^{-\kappa r_{\mathrm{p}} x} \left( \kappa r_{\mathrm{p}} + \frac{1}{x} \right) + \frac{3a_{2\mathrm{s}}}{x^3} \right) \ .
\end{equation}
For example, $f_{\mathrm{pp}} \approx 0.1$ pN at $r = 10 \ \mathrm{\mu m}$, which compares well to the experimental measurements in Figure 1 of Ref.~\cite{Masschaele2010}. For the contribution of interparticle interactions to the surface  pressure, we can use equation (\ref{eq:pi_from_du}) to write:
\begin{equation}
    \Pi_{U} = \frac{k_{\mathrm{B}} T}{r_{\mathrm{p}}^2 \sqrt{3}} \frac{1}{x^2} \left( \frac{a_{1\mathrm{s}}}{3} \mathrm{e}^{-\kappa r_{\mathrm{p}} x} \left( \kappa r_{\mathrm{p}} + \frac{1}{x} \right) + \frac{3a_{2\mathrm{s}}}{x^3} \right) \ .
\end{equation}
Hence, even at $r = 2.008 r_{\mathrm{p}}$, $\Pi_{U} \sim 0.005$ mN/m for these charged 3.1 $\mu$m diameter polystyrene particles at a water-oil interface. Notably, this suggests that extracting typical colloid pair potentials from  Langmuir-trough measurements does not seem feasible.

\section{Results}

\begin{figure}
    \centering
    \includegraphics[width=0.99\linewidth]{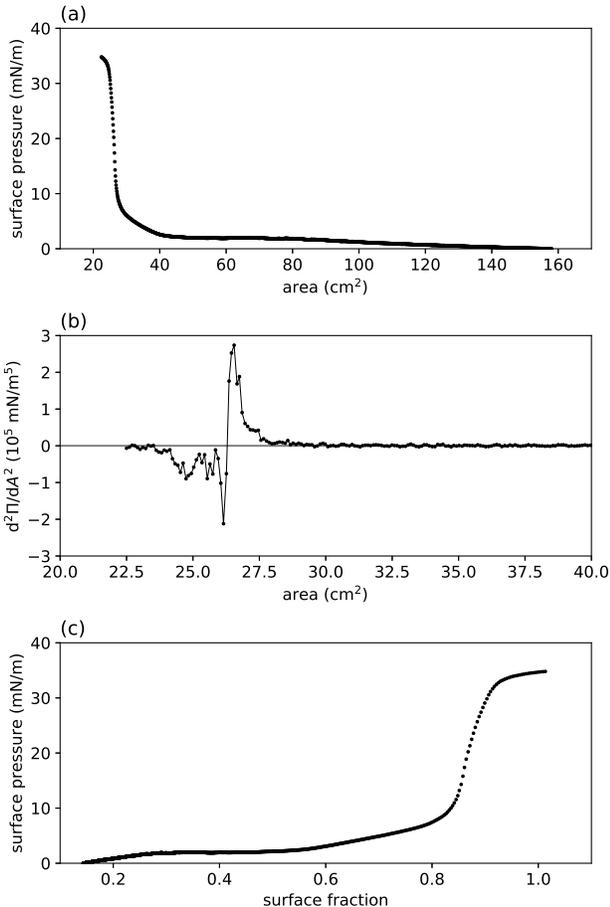}
    \caption{Compression measurements performed in a Langmuir trough for (undried) $0.455 \ \mu\mathrm{m}$ radius PMMA-PHSA particles at a water-hexadecane interface \cite{VanHooghten2017,VanHooghten2018data}. (a) Measured surface pressure $\Pi$ vs controlled area available to the interfacial particles. (b) Second derivative of $\Pi$, determined numerically from (a), to pinpoint the area at the inflexion point $A_{\mathrm{i}}$. The graph was smoothed by boxcar averaging to clarify where it crosses the horizontal axis ($A_{\mathrm{i}} \approx 26.3 \ \mathrm{cm}^2$). The solid line is a guide to the eye. (c) $\Pi$ vs surface fraction $\phi$, extracted from (a) by setting $\phi \left( A_{\mathrm{i}} \right) = 0.863$ \cite{Quickenden1974}.} 
    \label{fig:Figure1}
\end{figure}

We apply our theoretical framework to previously reported Langmuir-trough measurements for PMMA particles, stabilized by PHSA, at a water-hexadecane interface (Figure \ref{fig:Figure1}(a)) \cite{VanHooghten2018data}. As expected for particles that have been reported to behave as near-hard spheres in oil \cite{Bosma2002,Bryant2002}, the surface pressure is practically 0 at relatively large area. At intermediate area, the surface pressure is finite but small, which has previously been attributed to long-range interparticle interactions. At low area, the surface pressure 1) rises steeply (presumably because the particles start touching) and 2) levels off as the particle-laden interface starts buckling \cite{Aveyard2000}. However, plotting surface pressure vs area available to the interfacial particles is not always useful, as there is no guarantee that all particles added to the system make it to the interface, thereby making it challenging to compare Langmuir-trough measurements to other methods and/or between particle-interface combinations.

To allow comparison with other measurements, we convert area into surface fraction $\phi$ i.e.~the area covered by all the particles as a fraction of the total area available to the particles, for which at least one value of area is needed at which the value of surface fraction is known. Figure \ref{fig:Figure1}(b) shows the second derivative of the surface pressure vs area graph in Figure \ref{fig:Figure1}(a); the inflexion point $A_{\mathrm{i}}$ of the latter is where the second derivative crosses the horizontal axis. We assume that the inflexion point corresponds to the steep increase in coordination number of interfacial particles, where the surface fraction $\phi = 0.863$ \cite{Quickenden1974}; note that this does not seem too dissimilar from the procedure in, for example, Ref.~\cite{Maestro2015}. Figure \ref{fig:Figure1}(c) shows the graph of surface pressure vs surface fraction that corresponds to Figure \ref{fig:Figure1}(a). Note that the surface pressure levels off above $\phi \approx 0.9$, which aligns with the maximum surface fraction of interfacial disks in 2D being at $\phi_{\mathrm{c}} \approx 0.906$.

\begin{figure}
    \centering
    \includegraphics[width=0.99\linewidth]{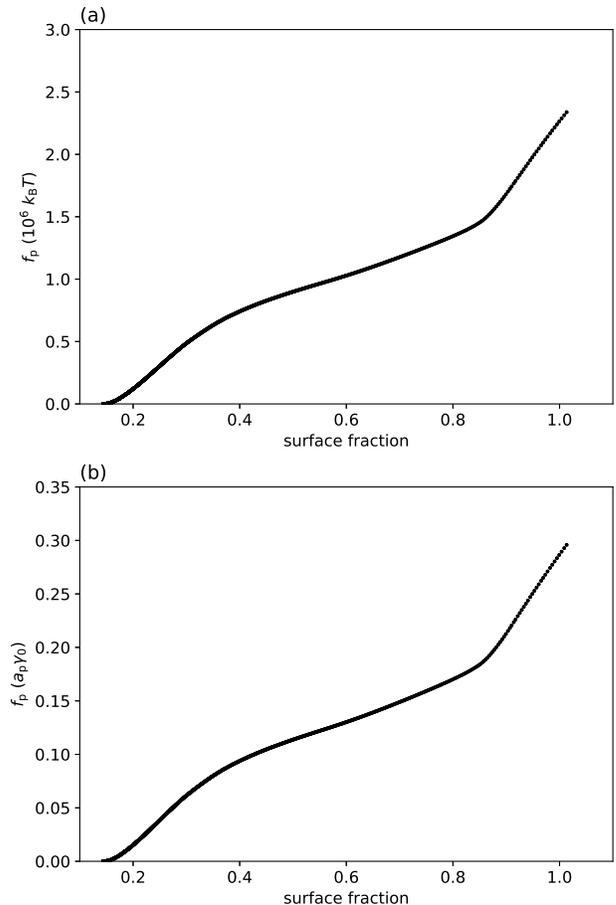}
    \caption{Free energy per particle vs surface fraction $\phi$: (a) $f_{\mathrm{p}}$ in units of $10^6 \ k_{\mathrm{B}} T$ and (b) $f_{\mathrm{p}}$ in units of $a_{\mathrm{p}}\gamma_0$. These graphs were extracted from the data presented in Figure \ref{fig:Figure1} using equation (\ref{eq:free_energy_particle_phi}).}
    \label{fig:Figure2}
\end{figure}

Following equation (\ref{eq:free_energy_particle_phi}), we numerically integrate the data in Figure \ref{fig:Figure1}(c) to obtain the free energy per particle due to the particles interacting, as $\bar{f}_{\mathrm{p}}$ is the quantity most closely related to the interparticle potential that we can extract from the data without further assumptions in our theoretical framework. Figure \ref{fig:Figure2}(a) shows $\bar{f}_{\mathrm{p}}$ as a function of surface fraction $\phi$. Note that, even for moderate values of surface fraction, where surface pressure is well below 5 mN/m, $\bar{f}_{\mathrm{p}}$ is of order $10^6$ $k_{\mathrm{B}}T$ i.e.~well beyond typical values for most colloidal interactions. In fact, plotting $\bar{f}_{\mathrm{p}}$ in units of $a_{\mathrm{p}} \gamma_0$ (Figure \ref{fig:Figure2}(b)) implies that interactions related to deformations of the liquid interface are at play here. In principle, these could be flotation capillary interactions \cite{Kralchevsky1994}, but the Bond number for these particles is $Bo \sim 10^{-8} << 1$ i.e.~flotation capillary forces are unlikely to be relevant here. Having said that, capillary forces caused by undulations of the contact line around the interfacial particles, e.g.~due to uneven stabilizer coverage, could play a role. However, we would expect these to lead to attractive interactions between the particles, whereas the surface pressure is positive (Figure \ref{fig:Figure1}), which points to repulsive interparticle interactions (equation (\ref{eq:hex_surface_pressure_ubar})); we will return to this discussion below.

\begin{figure}
    \centering
    \includegraphics[width=0.99\linewidth]{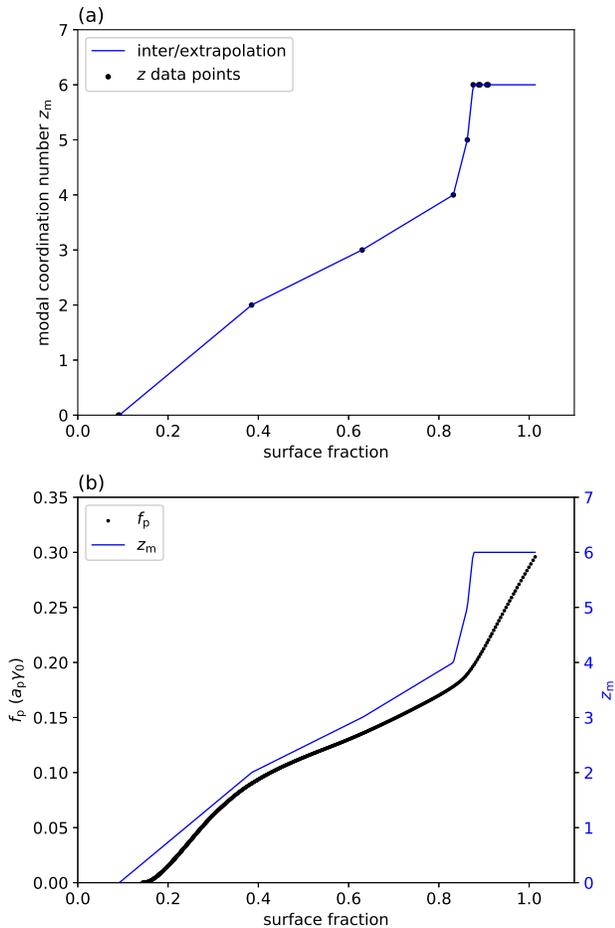}
    \caption{(a) Modal coordination number $z_{\mathrm{m}}$ of disks on an elastic sheet vs surface fraction $\phi$: solid circles are data points from Ref.~\cite{Quickenden1974} and the solid line is a linear interpolation (apart from $\phi > 0.9$ where we have set $z = 6$). (b) Combined graph of the free energy per particle from Fig.~\ref{fig:Figure2}(b) (bottom line) and the modal coordination number from panel (a); especially for $\phi \lesssim 0.83$, the shapes of the two graphs are remarkably similar.}
    \label{fig:Figure3}
\end{figure}

Even if liquid deformations could explain the order of magnitude for $\bar{f}_{\mathrm{p}}$, it is not immediately clear how they could explain the shape of the graphs in Figure \ref{fig:Figure2}. To better understand that shape, we take experimentally determined values of the modal coordination number $z_{\mathrm{m}}$ of (macroscopic) hard disks on an elastic sheet from Quickenden \emph{et al}.~ \cite{Quickenden1974} and plot them as a function of the surface fraction of the disks (Figure \ref{fig:Figure3}(a)). We interpolate between the available data points and we extropolate $z=6$ for $\phi > 0.9$, as $z = 6$ is the maximum coordination number of (hexagonally) close-packed disks in 2D. Intriguingly, the shape of the $\left( \bar{f}_{\mathrm{p}} , \phi \right)$-graph is described remarkably well by the shape of the $\left( z , \phi \right)$-graph, especially for $\phi \lesssim 0.83$ (Figure \ref{fig:Figure3}(b)). This suggests that the surface-pressure behaviour at low to intermediate surface fraction can be explained by the number of particle-particle contacts. Around $\phi = 0.83$, the modal coordination number $z_{\mathrm{m}}$ rises rapidly from 4 to 6, whereas $\bar{f}_{\mathrm{p}}$ rises less rapidly in that regime. One explanation could be that particle rearrangements due to interparticle interactions may affect $\left( z , \phi \right)$, especially at high surface fraction, which is not captured by the model system of hard disks on an elastic sheet. At even higher surface fractions, surface-pressure changes can no longer be explained by changes in coordination number, as $z_{\mathrm{max}} = 6$ has been reached.

\begin{figure}
    \centering
    \includegraphics[width=0.99\linewidth]{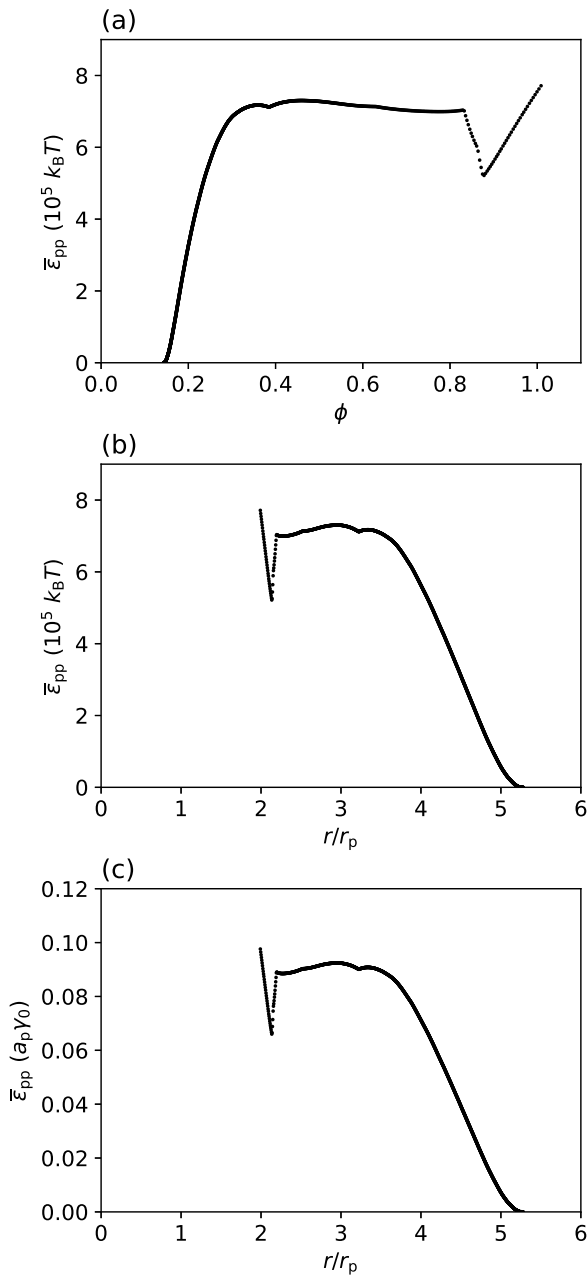}
    \caption{(a) Interparticle potential $\bar{\epsilon}_{\mathrm{pp}}$ vs surface fraction $\phi$. (b) $\bar{\epsilon}_{\mathrm{pp}}$ in units of $10^5 \ k_{\mathrm{B}} T$ vs separation $r$ in units of particle radius $r_{\mathrm{p}}$. (c) $\bar{\epsilon}_{\mathrm{pp}}$ in units of $a_{\mathrm{p}}\gamma_0$ vs $r / r_{\mathrm{p}}$. These graphs were extracted from the data presented in Figure \ref{fig:Figure3} using equation (\ref{eq:interparticle_potential_langmuir_trough}).}
    \label{fig:Figure4}
\end{figure}

If we assume that particles only interact with their nearest neighbours, and that the contribution of entropy to the surface pressure is negligible for micron-sized particles, then we can attemp to extract the interparticle potential $\bar{\epsilon}_{\mathrm{pp}}$ from surface-pressure measurements (equation (\ref{eq:interparticle_potential_langmuir_trough})). Figure \ref{fig:Figure4} shows the corresponding $\left( \bar{\epsilon}_{\mathrm{pp}} , \phi \right)$-graph and $\left( \bar{\epsilon}_{\mathrm{pp}} , r/r_{\mathrm{p}} \right)$-graphs, where the conversion from $\phi$ to the particle-particle separation $r$ has been done using $a \approx \pi \left( r/2 \right)^2$ and $\phi = a_{\mathrm{p}} / a$. As expected, the interparticle potential is negligible at large separations; it starts to increase around $r = 5 r_{\mathrm{p}}$ i.e.~$\phi \approx 0.16$. It then rises to a plateau value for $r \lesssim 3.5 r_{\mathrm{p}}$, corresponding to $\phi \gtrsim 0.33$. The height of this plateau, at approximately $0.09 a_{\mathrm{p}} \gamma_0$ or $7 \cdot 10^5 \ k_{\mathrm{B}}T$, supports the idea that deformations of the liquid interface are involved, as the free energy associated with the deformation of a liquid interface is expected to be of the order of the interfacial tension times the deformed area. Approaching close-packing, i.e.~near $r = 2 r_{\mathrm{p}}$, the interparticle potential features an unexpected dip. However, we attribute this to artefacts of the analysis. For example, given the steepness of the $\left( z, \phi \right)$ graph (Figure \ref{fig:Figure3}(a)), small differences in the $\left( z, \phi \right)$ behavior between disks on an elastic sheet and PMMA particles at a liquid interface can cause abrupt changes in $\bar{\epsilon}_{\mathrm{pp}} \left( r \right)$. Moreover, near close packing, the particles are close to jamming, at which point the interfacial particles are no longer in equilibrium and our thermodynamic approach breaks down. Finally, the particle-laden interface starts buckling for $\phi \gtrsim 0.9$ i.e.~$r \lesssim 2.1 r_{\mathrm{p}}$, which has not been taken into account in this analysis.

One might argue that the strong repulsion between the particles is due the their steric stabilization. However, a repulsive interaction of order $10^5 \ \mathrm{to} \ 10^6 \ k_{\mathrm{B}}T$ is beyond the measured repulsive barrier of sterically stabilized PMMA-PHSA particles \cite{Doroszkowski1971,Iacovella2010}. It is perhaps also surprising that the interparticle potential has exceeded $10^5 \ k_{\mathrm{B}}T$ at a relatively large separation of $r \approx 4.5 r_{\mathrm{p}}$. However, it should be noted that, especially at low surface fraction, the surface coverage is not necessarily homogeneous. For example, we have observed that buckling tends to start at the barriers rather than uniformly across the Langmuir trough \cite{Mears2020PhD}. Secondly, there may be a non-uniform stress distribution across the interface i.e.~a Janssen effect \cite{Cicuta2009}. Moreover, the barriers are typically moved at speeds and over distances that result in relatively high strain rates and total strains, for which careful constitutive modelling is required \cite{Verwijlen2014,Pepicelli2017}. All the same, our main claims so far are that i) measured surface pressures should be negligible for low surface fractions and ii) surface-pressure measurements cannot be used to extract typical colloid potentials; these claims are unaffected by these considerations.

Instead, we argue that the following picture emerges for the surface pressure of liquid interfaces laden with micron-sized particles. At very low surface fraction, i.e. when the interfacial particles are not interacting, the surface pressure is practically negligible. At low and intermediate surface fraction, the shape of the $\left( \Pi , \phi \right)$-graph can be explained by the particle coordination number i.e. the number of nearest neighbours of an interfacial particle. At high surface fraction, the surface pressure plateaus, which we attribute to buckling of the particle-laden liquid interface, in line with previous reports \cite{Aveyard2000,Thijssen2018}. The order of magnitude of the free energy per particle, and of the repulsive interparticle potential extracted from surface-pressure measurements, suggests that deformations of the liquid interface at the length scale of the particles are involved. We suggest that
these deformations are due to interfacial particles touching: given variance in particle size and contact angle \cite{Isa2011}, particle-particle contact forces will have components in the direction perpendicular to the liquid interface, leading to particles being pushed slightly out of the plane of the liquid interface. The length scale of these deformations will be of the order of the particle radius, so the free-energy cost per particle should indeed be of order $a_{\mathrm{p}} \gamma_0$. Alternatively, it could lead to the contact line of the liquid interface sliding along the interfacial particle, but it has previously been shown that this leads to free-energy changes of a similar order of magnitude (see SI of \cite{Maurice2013}).

\section{Conclusions}

We have presented here a theoretical framework to understand the surface pressure of liquid interfaces laden with micron-sized particles. As the particle detachment energy is several orders of magnitude larger than $k_{\mathrm{B}} T$, and hence the particles at the liquid interface are not in chemical equilibrium with those in the bulk, we derive an expression for the surface pressure in the canonical rather than the grand-canonical ensemble. We show that the surface pressure of a (dilute) collection of particles, with hard-core repulsion only, at a liquid interface is practically negligible (and actually zero if the entropy of the particles is ignored). Moreover, typical colloidal interactions, specifically those well below $10^5 \ k_{\mathrm{B}}T$, lead to surface pressures that are small to negligible on the scale of typical (measured) surface pressures. Instead, we argue that the shape of surface pressure-surface fraction graphs can be explained by particle coordination number at low to intermediate surface fractions; the order of magnitude of the free energy per particle extracted from surface-pressure measurements suggests that contact-line sliding and/or deformations of the liquid interface at the length scale of the particles play a pivotal role.

It is perhaps interesting to note that the system under consideration here could be considered as a 2D equivalent of the system studied by Guy \emph{et al}.~\cite{Guy2016}. They study the role of friction in the rheology of 3D suspensions of PMMA particles. In the system considered here, the system is Brownian at low surface fraction (see figure 1 in \cite{VanHooghten2017}). At high surface fraction, i.e.~when the surface pressure deviates substantially from 0, we have argued that contact forces start dominating the surface pressure, at which point the system is no longer Brownian. Hence, it would be interesting to consider what the role of friction is in Langmuir-trough experiments and our considerations here have provided an ansatz for that line of inquiry.

\section{Acknowledgements}

RM thanks EPSRC and IM thanks EPSRC and the CM-CDT for a PhD studentship (EP/M506515/1 and EP/L015110/1 respectively). JHJT thanks The University of Edinburgh for a Chancellor's Fellowship. The authors acknowledge Mike Cates, Willem Boon and Ren\'{e} van Roij for useful discussions.

\bibliography{ms}

\end{document}


\title{Surface Pressure of Liquid Interfaces Laden with Micron-Sized Particles - Supplementary Information}

\author{R. Mears, I. Muntz and J.~H.~J.~Thijssen}
\email[]{j.h.j.thijssen@ed.ac.uk}
\homepage[]{\\ www.ph.ed.ac.uk/people/job-thijssen}
\affiliation{SUPA School of Physics \& Astronomy, The University of Edinburgh, Edinburgh, EH9 3FD, Scotland, United Kingdom}

\date{\today}

\maketitle

\section{Theory}

\subsection{Surface pressure: derivation in canonical ensemble}

In the main text, we obtained an expression for the surface pressure $\Pi$, in two dimensions, via an analogy with the osmotic pressure of a suspension in three dimensions. Here, we derive an expression for the surface pressure of a particle-laden interface in the canonical ensemble, starting from the (canonical) free energy $F$ of the particle-laden interface.

In analogy with the derivation for the surface tension of surfactant solutions \cite{Doi2015Ch4}, we start with the following expression for particles at liquid interfaces:
\begin{equation}{\label{eq:can_interfacial_energy_init}}
F \left( A, T, N \right) = A f \left( T, N \right) \ ,
\end{equation}
where $F$ is the (canonical) free energy of the interface, $A$ the area of the interface available to the particles, $T$ the temperature and $N$ the number of interfacial particles. Note that $f \left( T, N \right)$ is a free-energy density, i.e.~free energy per unit area, so it is counter-intuitive that it depends on the total number of interfacial particles $N$. This inconsistency stems from $N$ being an extensive variable, whereas $T$ and the chemical potential $\mu$ (the variables of the free-energy density in the grand-canonical ensemble) are both intensive variables. An intensive variable related to $N$ is the particle number density:
\begin{equation}{\label{eq:number_density}}
\rho = \frac{N}{A} \ ,
\end{equation}
so we suggest:
\begin{equation}{\label{eq:can_interfacial_energy_final}}
F \left( A, T, N \right) = A f \left( T, \rho \right) \ .
\end{equation}

For the full differential of the canonical free energy, we can write \cite{Kittel1998, Doi2015Ch4}:
\begin{equation}{\label{eq:can_energy_differential_alt}}
\mathrm{d} F = \gamma \mathrm{d}A -S \mathrm{d}T + \mu \mathrm{d}N \ .
\end{equation}
From equation (\ref{eq:can_interfacial_energy_final}), we obtain:
\begin{equation}{\label{eq:can_energy_differential}}
\begin{array}{rcl}
\mathrm{d} F & = & f \mathrm{d} A + A \mathrm{d}f \\
\vspace{0.5mm} \\
& = & f \mathrm{d}A + A \left( \frac{\partial f}{\partial T} \right)_{A,\rho} \mathrm{d}T + \\
& + & A \left( \frac{\partial f}{\partial \rho} \right)_{A,T} \mathrm{d}\rho \ .
\end{array}
\end{equation}
To compare terms in equations (\ref{eq:can_energy_differential_alt}) and (\ref{eq:can_energy_differential}) like for like, we have to write $\mathrm{d}\rho$ in terms of $\mathrm{d}N$ and $\mathrm{d}A$:
\begin{equation}{\label{eq:can_density_differential}}
\begin{array}{rcl}
\mathrm{d} \rho & = & \left( \frac{\partial \rho}{\partial N} \right)_{A, T} \mathrm{d}N + \left( \frac{\partial \rho}{\partial A} \right)_{N, T} \mathrm{d}A \\
\vspace{0.5mm} \\
& = & \frac{\mathrm{d}N}{A} - \frac{\rho \mathrm{d}A}{A} \ .
\end{array}
\end{equation}
Substituting equation (\ref{eq:can_density_differential}) into equation (\ref{eq:can_energy_differential}), we get:
\begin{equation}{\label{eq:can_energy_differential_final}}
\begin{array}{rcl}
\mathrm{d} F & = & f \mathrm{d}A + A \left( \frac{\partial f}{\partial T} \right)_{A,\rho} \mathrm{d}T \\
& + & A \left( \frac{\partial f}{\partial \rho} \right)_{A,T} \left( \frac{\mathrm{d}N}{A} - \frac{\rho \mathrm{d}A}{A} \right) \\
\vspace{0.5mm} \\
& = & \left( f - \rho \left( \frac{\partial f}{\partial \rho} \right)_{A,T} \right) \mathrm{d}A + A \left( \frac{\partial f}{\partial T} \right)_{A,\rho} \mathrm{d}T \\
& + & \left( \frac{\partial f}{\partial \rho} \right)_{A,T} \mathrm{d}N \ .
\end{array}
\end{equation}

Now, we can compare terms like for like between equations (\ref{eq:can_energy_differential_alt}) and (\ref{eq:can_energy_differential_final}), resulting in:
\begin{equation}{\label{eq:can_differential_equations}}
\begin{array}{rcl}
\left( \frac{\partial f}{\partial T} \right)_{A, \rho} & = & -\frac{S}{A} \\
\vspace{0.5mm} \\
\left( \frac{\partial f}{\partial \rho} \right)_{A, T} & = & \mu = \left( \frac{\partial F}{\partial N} \right)_{A, T} \\
\vspace{0.5mm} \\
\gamma & = & f - \rho \left( \frac{\partial f}{\partial \rho} \right)_{A,T} ,
\end{array}
\end{equation}
where the right-hand side of the second row employs the definition of chemical potential in the canonical ensemble \cite{Kittel1998}. Combining rows 2 and 3 of equation (\ref{eq:can_differential_equations}), we get an expression for the effective interfacial tension $\gamma$ of a colloid-laden interface in the canonical ensemble:
\begin{equation}{\label{eq:can_gamma}}
\gamma = f - \rho \mu \ .
\end{equation}
Typically, we are interested in the surface pressure of the colloid-laden interface, which in the canonical ensemble is then:
\begin{equation}{\label{eq:can_surface_pressure}}
\Pi = \gamma_0 - \gamma = \gamma_0 - f + \rho \mu \ .
\end{equation}
Note that equation (\ref{eq:can_surface_pressure}) here is equivalent to equation (6) in the main text.

\bibliography{supplement}